\documentclass[journal=jacsat,manuscript=article]{achemso}

\usepackage[version=3]{mhchem} 
\usepackage[x11names]{xcolor}  
\usepackage{courier}  
\usepackage[T1]{fontenc}  
\usepackage{amsmath}  
\usepackage{amssymb}  
\usepackage{amsfonts}  
\usepackage{mciteplus} 
\usepackage{graphicx}  
\usepackage{achemso}
\usepackage{hyperref}
\usepackage{titlesec}

\usepackage[margin=10pt,font=small,labelsep=period]{caption} [2023/03/12]
\usepackage{stackengine} 
\usepackage{booktabs,siunitx}
\usepackage{nicematrix,tikz}  
\usepackage{outlines} 
\usepackage{enumitem} 
\usepackage{comment}
\usepackage{pdfpages}

\setenumerate[1]{label=\Roman*.} 
\setenumerate[2]{label=\Alph*.}
\setenumerate[3]{label=\roman*.}
\setenumerate[4]{label=\alph*.}
\author{Kuang-Hua Chou}
\author{Alexander Eden}
\author{David E. Huber}
\author{Sumita Pennathur}
\affiliation[UCSB MechE]
{Department of Mechanical Engineering, University of California, Santa Barbara}
\author{Deborah~Kuchnir~Fygenson}
\email{fygenson@ucsb.edu}
\affiliation[UCSB Physics]
{Department of Physics, University of California, Santa Barbara}

\title[Nanostars in nanochannels]
  {Electrokinetic nanofluidic sensing of DNA nanostar condensate}

\abbreviations{DNA,NaCl}
\keywords{current monitoring, phase transition, critical temperature}

\begin{document}

\begin{abstract}
We demonstrate electronic sensing of DNA nanostar (NS) condensate.    
Specifically, we use electrokinetic nanofluidics to observe and interpret how temperature-induced NS condensation affects nanochannel current. 
The increase in current upon filling a nanochannel with NS condensate indicates that its electrophoretic mobility is about half that of a single NS and its effective ionic strength is $\sim35$\% greater than that of 150mM \ce{NaCl} in phosphate buffer.
$\zeta$-potential measurements before and after exposure to NS show that condensate binds the silica walls of a nanochannel more strongly than individual NS do under identical conditions.
This binding increases electroosmotic flow, possibly enough to completely balance, or even exceed, the electrophoretic velocity of NS condensate.
Although the current through a flat nanochannel is erratic in the presence of NS condensate, tilting the nanochannel to accumulate NS condensate at one entrance (and away from the other) results in a robust electronic signature of the NS phase transition at temperatures $T_c = f([\ce{NaCl}])$ that agree with those obtained by other methods.
Electrokinetic nanofluidic detection and measurement of NS condensate thus provides a foundation for novel biosensing technologies based on liquid-liquid phase separation. 
\end{abstract}

\newpage
\section{Introduction}
Nucleic-acid nanostars (NS) are programmable supramolecules with applications in fields ranging from material science to cell biology due to their ability to undergo liquid-liquid phase separation (LLPS). 
In buffered saline, a homogeneous solution of NS separates into a dense (NS-rich) aqueous phase in equilibrium with a dilute (NS-poor) aqueous phase upon cooling~\cite{biffi2013phase}.
The dense phase, also known as `NS condensate', is in many ways analogous to naturally occurring biomolecular condensates –– membraneless organelles that regulate transcription~\cite{transcriptional_condensates_2024} and control cytoplasmic availability of small molecules or proteins in cells~\cite{MLO_review_2023}.
Accordingly, NS condensates are being engineered to serve as model biomolecular condensates~\cite{agarwal_dynamic_2024,fabrini_co-transcriptional_2024,maruyama_temporally_2024, stewart_modular_2024}, expanding the range of their potential application to include cellular engineering, biogenesis, synthetic biology, and biotechnology~\cite{malouf2023sculpting, OOL_LLPS_2021, synbio_LLPS_2024}. 
There is therefore great interest in gaining a fundamental understanding of NS LLPS and how to control it via environmental conditions ({\em e.g.}, salt, pH, and temperature)~\cite{rovigatti2014accurate, conrad2022emulsion} and macromolecular design ({\em e.g.}, composition, sequences, and structure)~\cite{rovigatti2014gels, Flex-based_2021}.
Systematic studies are hampered, however, by the cost and complexity involved in NS condensate detection and characterization.

To date, with the notable exception of bulk rheometry~\cite{conradPNAS_2019}, NS condensates are exclusively detected optically.
Techniques such as light scattering~\cite{bomboi2015equilibrium, biffi2013phase} and fluorescence microscopy~\cite{sato2020sequence, agarwal2022growth, conrad2022emulsion} offer excellent sensitivity, but require large quantities, chemical modifications and/or a high degree of purification, making high throughput prohibitively expensive. 
The ability to detect condensate electrically, instead of optically, could remove these constraints and thereby facilitate characterization of a greater variety of NS designs over a broader range of conditions.   

Here we demonstrate, for the first time, electrical detection of NS condensate. 
Our approach is based on electrokinetic nanofluidics, in which fluid is driven through a nanoscale channel by an applied voltage and its flow is detected from changes in the resulting electrical current.
It is well established that the high surface area to volume ratio, confined geometry and finite electric double layer in nanochannels can be used to separate, identify, and study DNA~\cite{branton2008potential,persson2010dna,stein2010electrokinetic,dorfman2010dna,liu2016slowing,nouri2021nanofluidic}.
Nanochannel conductivity, in particular, has been used to detect single-stranded DNA (ssDNA) and double-stranded DNA (dsDNA) in a rapid, sensitive and versatile manner~\cite{crisalli2015label}. 

The present work lays a foundation for the rapid, sensitive, and versatile electrical detection and characterization of NS condensate. 
Using four-armed NS made of minimally purified synthetic DNA, we determine that NS condensate is more conductive than the uniform phase from which it condenses.
We characterize its electrophoretic mobility and conductivity by filling a nanochannel with NS condensate and measuring the associated increase in nanochannel current.
We go on to assess the change in $\zeta$-potential inside a nanochannel that results from NS condensate binding to the silica walls.
We find this change is capable of limiting the amount of NS condensate that enters the nanochannel and proceed to interpret the electrical signatures of NS LLPS in terms of known electrokinetic effects. 
Finally, we identify the NS LLPS transition temperature and show that the presence of NS condensate from as little as 20~$\mu$L of 50~$\mu$M NS in a background of anywhere from 50~mM to 300~mM \ce{NaCl} can be robustly detected.

\section{Materials and methods}
All DNA strands were purchased with standard desalting (Integrated DNA Technologies) and used without further purification.  
All chemicals were of analytical grade and used without further purification. 
All solutions were prepared with highly purified water of $18.2$~M$\Omega/$cm resistivity (Milli-Q Gradient A10, Millipore-Sigma).

\subsection{Sample preparation} 
All samples were prepared in home-made phosphate-buffered saline solution (PBS) containing 19~mM phosphate (80\% \ce{Na_2HPO_4}, 20\% \ce{NaH_2PO_4}) with an additional 50, 150, or 300~mM \ce{NaCl}, which we refer to as PBS050, PBS150, and PBS300, respectively.
Their room temperature pH values were 7.4, 7.3 and 7.2, respectively, and varied $\leq 0.02$ in the range $20^\circ<T<60^\circ$. 

Two different, four-armed DNA nanostars (NS) were used, one capable of undergoing LLPS ($\alpha$-NS) and one incapable of doing so ($\tau$-NS) (Fig.~\ref{fgr:allSchematics}; Supp Info, Section 1).
Each consisted of four 49-base oligomers (Table~S1).
The sequences of $\alpha$-NS oligomers were the same as in prior works \cite{conrad2022emulsion, biffi2013phase}.
The sequences of $\tau$-NS oligomers were identical to those of $\alpha$-NS except in their 3$^\prime$-overhangs, where the six-base, self-complimentary ``$\alpha$'' sequence (\emph{i.e.}, ``sticky end'') was replaced with six (non-complimentary) thymine bases. 
Each oligomer was hydrated in pure water to a concentration of $1000~\mu$M in the manufacturer's tube by alternately submersing the sealed tube containing hydrated oligomer in a $60^\circ$C water bath for 5~min and vortexing for one minute, three times.
The concentration of each oligomer stock solution was determined from the absorbance of 260~nm light (Nanodrop 2000c, Fisher Scientific) by a 100-fold dilution in pure water using the sequence-specific extinction coefficient provided by the manufacturer (Table~S1).
Stock solutions were placed in low-retention tubes (cat no. 24-282LR, Genesee Scientific) for long-term storage at $-80^\circ$C or short-term storage (<1~wk) at $-20^\circ$C.

To prepare NS, oligomer stock solutions were combined to achieve a 1:1:1:1 stoichiometry in a low-retention tube, then wrapped in foil, placed in an upside-down dry block (StableTemp~125D, Cole-Parmer) and covered with a two-inch thick piece of styrofoam ({\em i.e.}, the lid of a shipping container).
The dry-block was heated to 90$^\circ$C and held at that temperature for 5~min.
Finally, power to the dry-block was turned off and the sample was left to cool slowly to room temperature over $\sim$12~hours to promote defect-free self-assembly of NS.

\subsection{Nanofluidic Device Fabrication} 
Nanochannels were {\color{black}custom} made from two fused silica wafers (model 4W55-325-15C, Hoya) using an optimized, room temperature bonding technique.~\cite{boden2017process} 
Briefly, two parallel trenches (5~mm~$\times$~5~$\mu$m~$\times$~100~nm), spaced 5~mm apart, were dry etched (E626I, Panasonic) in one silica wafer and four 1.5~mm diameter holes, centered on the corners of a 5~mm~$\times$~5~mm square, were drilled in the other.
Silanol (\ce{SiOH}) groups were created on the surfaces of both wafers by exposure to \ce{O2} and then \ce{N2} in a capacitively-coupled plasma (EVG810, EV Group) for 1-2 minutes.
The plasma treated surfaces were aligned and placed in contact with each other, creating irreversible \ce{Si-O-Si}, \ce{Si-O-N-Si}, and \ce{Si-N-N-Si} bonds~\cite{xu2018glass}.
The bonded wafer was diced into chips (model 7100, Advanced Dicing Technologies) without post-bond annealing~\cite{chou2018experimental}. 

\subsection{Experimental Setup} 
A chip was sandwiched between the two halves of a Delrin chip holder, the upper half of which created reservoirs ($9.4~\mu$L) above each drilled hole ($0.92~\mu$L) (Fig.~S1A). 
A thin (0.015 in, 10A Duro) silicone sheet (cat no. 86435K131, McMaster-Carr) served as a gasket between the top surface of the chip and the bottom of the upper half of the chip holder (Fig.~S2).
Two counter-sunk screws secured the top half of the chip holder to the bottom half, and compressed the gasket when tightened. 
Two pairs of $0.5$~mm OD platinum electrodes (World Precision Instruments) accessed the reservoirs through holes in the sides of the top half of the chip holder.
To minimize evaporation at high temperatures, the reservoirs were sealed with electrical tape (Super 33+, Scotch) and the electrode access holes were filled with high-vacuum grease (Dow Corning). 

Temperature control was established using one or two programmable baths (Corio CD-200F and 300F, Julabo) to either change temperature slowly and steadily or rapidly switch between two temperatures.
To achieve rapid switching, a pair of motorized 4-way valves (835-PL4F4F4F4F-N, Specialty Manufacturing), controlled by a microcomputer (model 4, Raspberry Pi), directed fluid from one bath or the other through insulated tubing and then through a metal platform (C203W, Biolin Scientific).
The chip holder was surrounded by copper blocks on top, bottom and one side such that copper provided a direct thermal connection between the aluminum platform and the bottom of the chip as well as between the platform and the electrical tape that sealed the top of the chip holder reservoirs.
Temperature was monitored by a thermocouple (6517-TP, Kiethley) situated in a hole directly below the chip in the lower copper block (Fig.~S2). 
The entire apparatus (silica chip, Delrin holder, copper blocks and aluminum platform) was placed in the bottom half of a styrofoam box within a well-insulated, metal-enclosed volume (Heratherm IGS100, Thermo Scientific) to minimize thermal fluctuations and electrical interference from the environment. 
Precision electrometers (6517A, Keithley) were used to simultaneously apply voltage ($\pm20$~V), measure current ($\sim$nA), and monitor temperature. 
The microcomputer and electrometers were controlled, and data was collected, by a desktop computer running MATLAB 2021b (Mathworks, Inc.).
The high resistance imposed by the geometry of the nanochannel ensured that 99\% of the applied voltage dropped across the nanochannel, creating a 4~V/mm electric field.

\section{Results \& Discussion}

\subsection{Electrical signatures of NS condensate}
\begin{figure}[h]
\includegraphics[width=15cm]{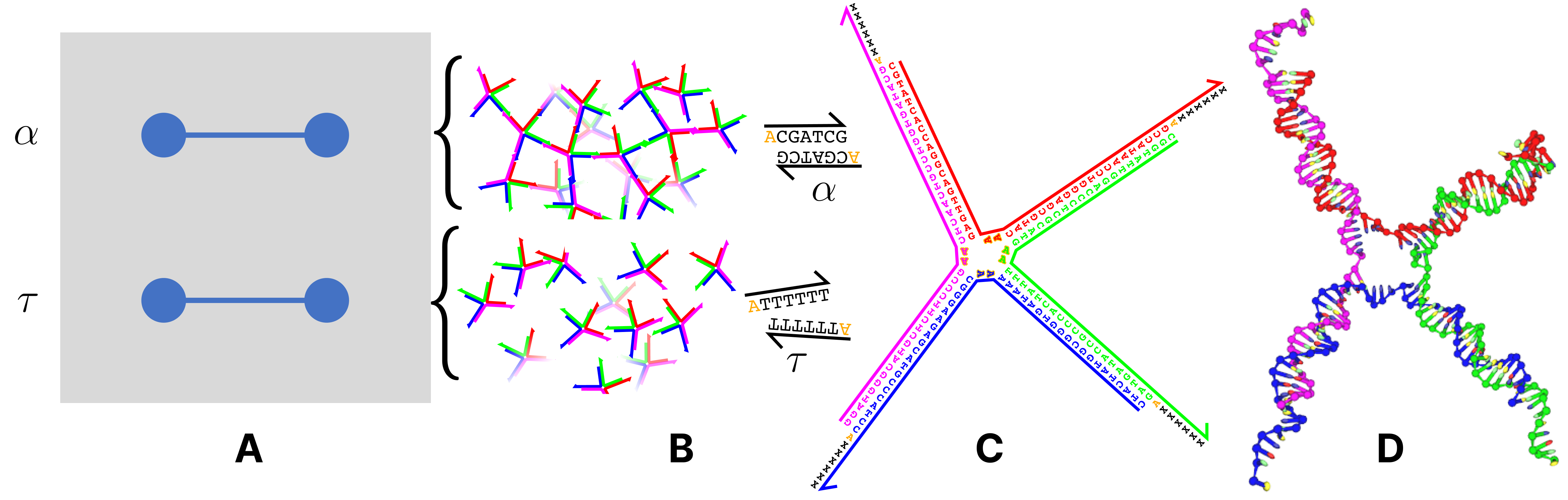}
\caption{Experimental designs. 
    {\bf\textsf{A}}:~Fused silica chip with two parallel 5~mm$\times$5~$\mu$m$\times$100~nm nanochannels, each flanked by 1~mm diameter access holes. 
    {\bf\textsf{B}}:~One channel is filled with a solution of $\alpha$-NS, whose self-complimentary overhangs can form NS-NS bonds; the other is filled with a solution of $\tau$-NS, whose non-complimentary overhangs cannot. 
    {\bf\textsf{C}}:~Strand sequence schematic and {\bf\textsf{D}}:~coarse-grained model~\cite{oxDNA,sulcOxView_2022} of the NS used in this study. 
    All four arms end in the same 6-base, single-stranded $3'$-overhang. 
    Unpaired adenines between arms and one at the base of the overhang provide flexibility at the NS junction and about the NS-NS bond \cite{NguyenSaleh2017}.
}
  \label{fgr:allSchematics}
\end{figure}

We looked for an electrical signature of DNA nanostar (NS) condensation in the temperature-dependent currents through a pair of 100~nm deep channels on a single fused silica chip (Fig.~\ref{fgr:allSchematics}A).
Both nanochannels were filled with a phosphate-buffered saline solution (PBS) containing four-armed NS (Fig.~\ref{fgr:allSchematics}B-D).
In one nanochannel, the $\tau$-channel, the NS had poly-T overhangs, which make them incapable of condensing. 
In the other nanochannel, the $\alpha$-channel, the NS had palindromic ({\em i.e.}, self-complimentary) overhangs known to mediate condensation below a critical temperature $T_c<50^\circ$C~\cite{conrad2022emulsion}.

\begin{figure}[h]
\includegraphics[width=7.5cm]{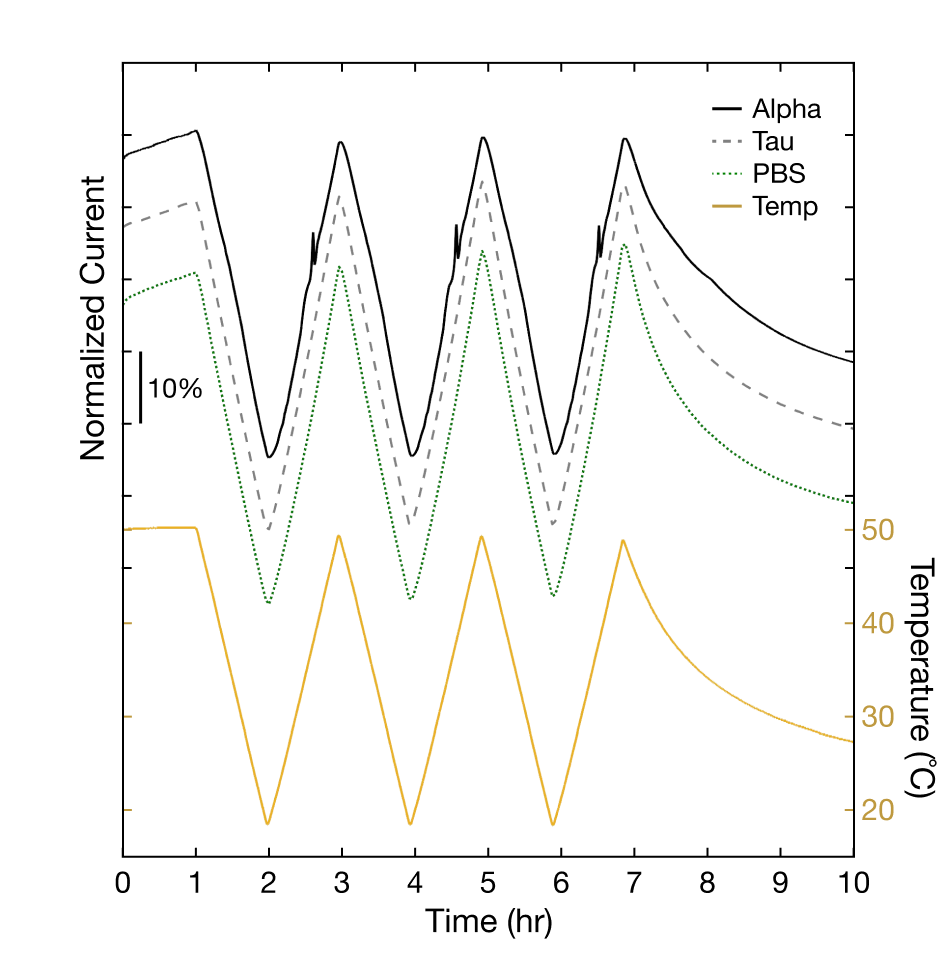}
\caption{
    Nanostar condensation affects nanochannel current.
    Current through a nanochannel (normalized to its highest value) rises and falls with temperature (solid, gold) as expected when it is filled with PBS150 alone (dotted, green) or 40~$\mu$M~$\tau$-NS in PBS150 (dashed, grey).  
    When it is filled with 40~$\mu$M~$\alpha$-NS in PBS150 (solid, black), the current through the nanochannel has irregularities that are repeatable from run to run but differ from prep to prep (data not shown).
}
  \label{fgr:detection}
\end{figure}

Steadily lowering and raising the temperature between $50^\circ$C and $20^\circ$C, at a rate of $0.5^\circ$C/min, had a negligible effect on the current through the $\tau$-channel (Fig.~\ref{fgr:detection}, dashed grey).  
It behaved exactly like current through a channel filled with PBS alone: increasing or decreasing with temperature as expected due to ionic mobility changing with the viscosity of water (Fig.~\ref{fgr:detection}, dotted green).
Current through the $\alpha$-channel behaved differently. 
It often spiked upon heating through a particular temperature but varied only slightly upon cooling back through the same temperature (Fig.~\ref{fgr:detection}, solid black).
The details of this behavior ({\em e.g.}, the specific temperature at which current spiked, the amplitude and width of the spike) were highly variable. 
Sometimes the oft-seen spike was completely absent (Fig.~S3).

To understand this signal and the conditions necessary for its robust and reproducible detection, we performed current monitoring (Supp Info, Section 2).
When filled with PBS alone ({\em i.e.}, without NS), the electroosmotic flow (EOF) through the nanochannel was directed toward the negative electrode (Table~\ref{tbl:EOF}).
This direction of flow is expected because negatively charged silica surfaces attract positive, mobile counterions (H$^+$ and Na$^+$), which drag fluid toward the negative electrode.
It is also opposite to the electrophoretic motion of DNA, whose negatively charged phosphodiester backbone is attracted toward the positive electrode.
Therefore, the speed and direction of individual NS and of NS condensates through the nanochannel depend on the magnitudes of their respective electrophoretic velocities (EPV) relative to the EOF. 

\begin{table}[hb]
  \begin{center}
      \begin{tabular}{|c|c|} 
        \hline
        {\color{black}Buffer} & EOF ($\mu$m/s)  \\
        \hline
         {\color{black}PBS300} & $37\pm2$  \\
        \hline
        {\color{black}PBS150} & $51\pm2$  \\
        \hline
        {\color{black}PBS050}  & $74\pm6$  \\
        \hline
   \end{tabular}
  \captionsetup{position=bottom}
   \caption{Electroosmotic flow {\color{black}of the three (NS-free) buffers used in this work (Sec. 2.1) through silica nanochannels (5~mm~$\times$~5~$\mu$m~$\times$~100~nm)}, as measured by current monitoring at room temperature.}
    \label{tbl:EOF}
\end{center}
\end{table}

\begin{figure}[h]
\includegraphics[width=7.5cm]{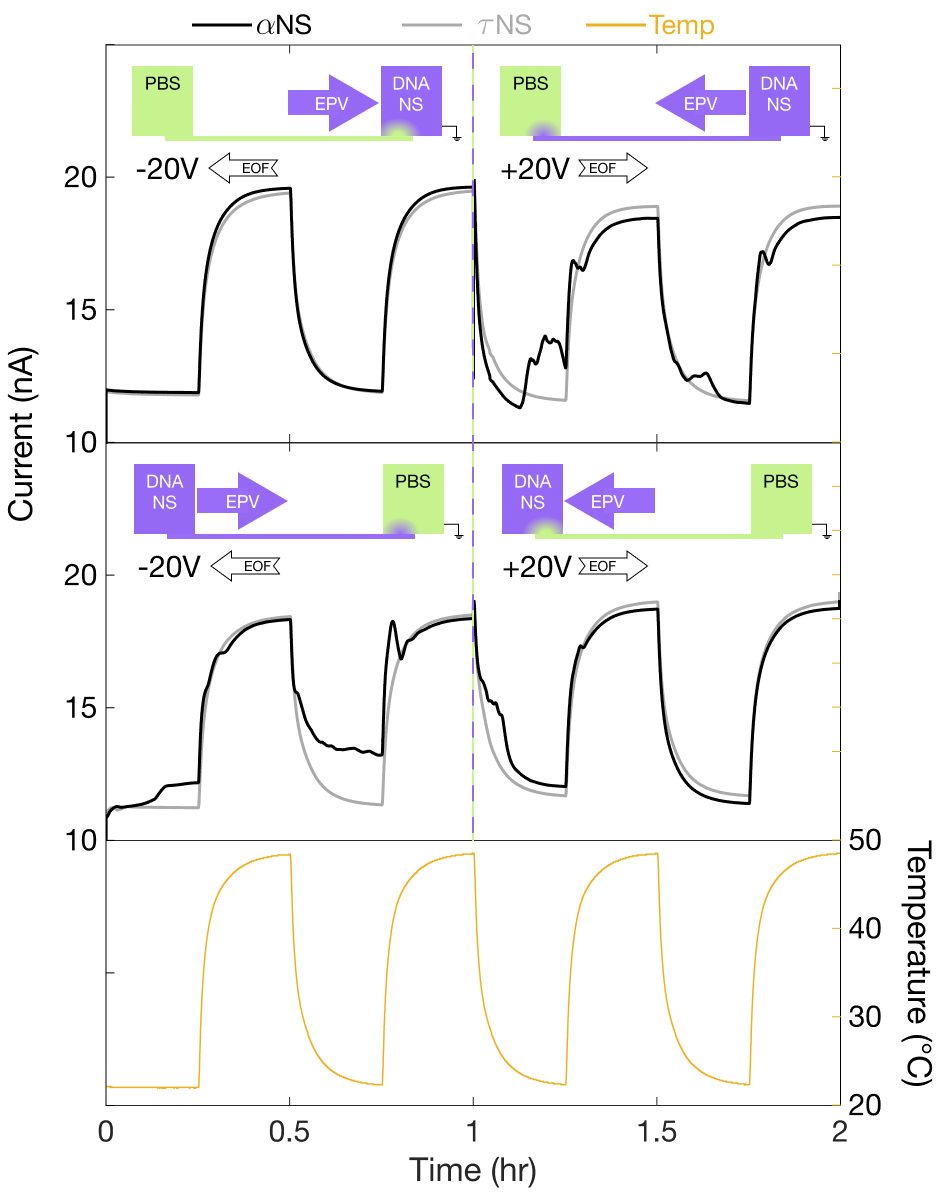}
\caption{
    Electrophoretic velocity (EPV) of NS condensate exceeds electroosmotic flow (EOF) of PBS150.
    When the reservoir at one end of a nanochannel is filled with PBS150 alone (green) and the reservoir at the other end is filled with 50$~\mu$M NS in PBS150 (purple), irregularities in nanochannel current occur at low temperatures only in the $\alpha$-channel, where NS are capable of condensing (thick black line).
    $\alpha$-channel current irregularities arise when EPV is directed into the nanochannel from the NS-laden reservoir and result in $\alpha$-channel currents that typically exceed $\tau$-channel current (thin grey line), suggesting that the presence of NS condensate makes the nanochannel more conductive than when it is filled with the homogeneous NS solution from which it phase separates. 
}
\label{fgr:EPvEOF}
\end{figure}

Despite their branched structure, individual NS have been shown~\cite{saha_ep_2006} to have the same, well-known electrophoretic mobility as dsDNA in solution~\cite{stellwagen2002determining}, $\mu_\text{dsDNA} \approx 3\cdot 10^4~\mu$m$^2$/V$\cdot$s. 
Under our conditions, this corresponds to an EPV~$=~\mu_\text{dsDNA}\times 4$~V/mm~$\approx120~\mu$m/s, which exceeds the EOF in our nanochannels at every salt concentration we explored (Table~\ref{tbl:EOF}).
The electrophoretic mobility of NS condensate, $\mu_\text{NSc}$, by contrast, is not known.
To get a sense for how the EOF compared to the EPV of NS condensate in our system, we filled both reservoirs with PBS150 and put 50~$\mu$M NS in only one of them.
We then monitored the nanochannel current while switching between high (50$^\circ$C) and low (20$^\circ$C) temperature under all possible polarities of the imposed voltage (Fig.~\ref{fgr:EPvEOF}).

When EPV was directed out of (and EOF into) the nanochannel at the base of the NS-laden reservoir, $\alpha$- and $\tau$-channel currents behaved similarly as temperature changed from cold to hot and back again  (Fig.~\ref{fgr:EPvEOF}, upper left and lower right).
When EPV was directed into (and EOF out of) the channel at the base of the NS-laden reservoir, $\tau$-channel current behaved as before but $\alpha$-channel current became irregular at low temperatures, where NS LLPS is favorable, and smooth at high temperatures, where it is not (Fig.~\ref{fgr:EPvEOF}, upper right and lower left).
This pattern suggests that, at least when EOF~$\approx50~\mu$m/s, NS condensate moves toward the anode (EPV > EOF) and enters the nanochannel.
Furthermore, the irregular $\alpha$-channel current often exceeded the $\tau$-channel current, suggesting that NS condensate is more conductive than uncondensed NS ({\em i.e.}, homogeneous phase).

\subsection{\label{ionicStrengthMeasurement}Measuring the ionic strength of NS condensate}
To determine how much more conductive the dense phase is and estimate its electrophoretic mobility, we sought to measure the proportional increase in current that occurs when NS condensate completely fills a nanochannel.
To achieve filling, we tilted the chip along the length of the nanochannels and allowed condensate to accumulate while EPV was directed uphill, then switched polarity.
We reasoned that 
({\em i}) because NS condensate sediments~\cite{jeon2018salt, wilken2023spatial}, tilting should cause it to pool away from the nanochannel entrance in the lower reservoir and toward the entrance in the upper reservoir; and 
({\em ii}) because the electric field is much stronger in the nanochannel than in the reservoirs, an uphill EPV might prevent NS condensate from entering the nanochannel from the upper reservoir while a downhill EOF, combined with gravity, might suffice to prevent condensate from approaching the $\alpha$-channel from the lower reservoir (Fig.~S1B).
Specifically, we filled and sealed both reservoirs with 50~$\mu$M~NS in PBS150 at a temperature well above the condensation temperature ($50^\circ$C), imposed an 11$^\circ$ tilt, applied +20~V to the upper reservoir and grounded the lower reservoir, then cooled to well below the condensation temperature ($20^\circ$C), and, after allowing NS condensate to accumulate for half an hour, switched polarity.

\begin{figure}[h]
    \begin{center}
        \includegraphics[width=10cm]{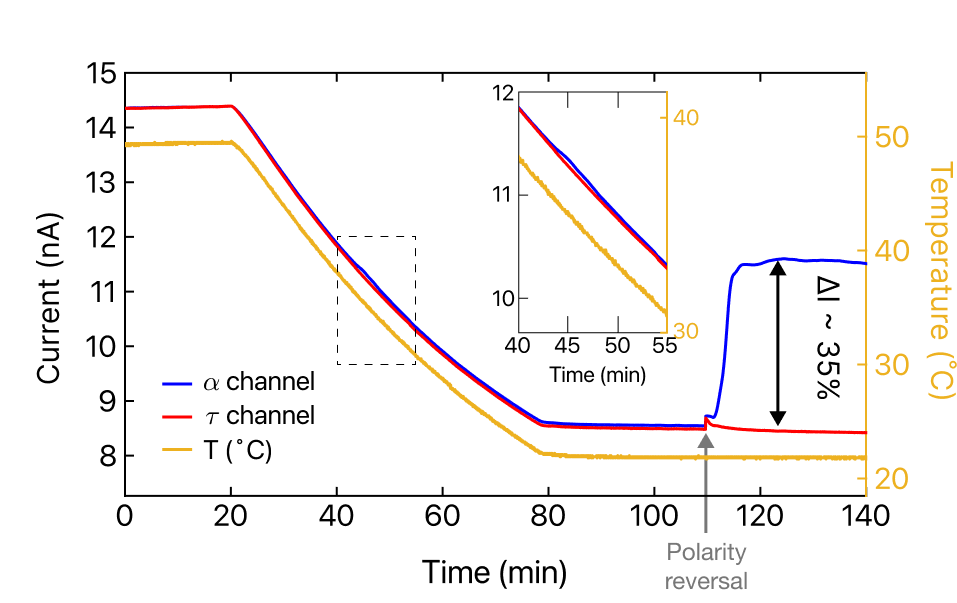}
    \end{center}
        \caption{NS condensate is more conductive than the uniform phase.
        Tilted $\alpha$- and $\tau$-nanochannels filled with 50~$\mu$M NS in PBS150 had similar currents at 50$^\circ$C and at 20$^\circ$C when EPV was directed uphill.
        While cooling, $\alpha$-channel current consistently deviated $\sim1\%$ relative to $\tau$-channel current (Inset).
        When polarity was reversed (after 20~min at 20$^\circ$C), current through the $\tau$-channel fluctuated briefly while the current through the $\alpha$-channel increased $\gtrsim 35\%$ and then held steady.
        Inset: Close-up of the slight deviation in $\alpha$-channel current that consistently occurred upon cooling.
        }
        \label{Figure measurement}
\end{figure}

This strategy was effective. 
While cooling with EPV directed uphill, the current decreased according to the changing viscosity of water in both channels (Fig.~\ref{Figure measurement}), with the remarkably consistent exception of a slight ($\sim 1\%$) variation in $\alpha$-channel current along the way (Fig.~\ref{Figure measurement}, inset).
While waiting for condensate to accumulate, both channel currents held steady. 
Finally, upon switching the voltage polarity to direct EPV downhill, the magnitude of the $\tau$-channel current fluctuated briefly, but the magnitude of the $\alpha$-channel current increased steadily for 3~min, then maintained a value $\gtrsim 35\%$ higher than the original for >20 min.
We conclude that $\alpha$-NS condensate filled the 5-mm-long nanochannel at a speed of $\sim30~\mu$m/sec and with an ionic strength $\gtrsim 200$~mM NaCl (Supp Info, Section 3). 

These observations have interesting implications.
Regarding condensate ionic strength:
Given that the concentration of DNA backbone phosphates in NS condensate is $\sim75$~mM ($196~\mathrm{bases/NS} \times 375~\mu$M~NS~in~condensate~\cite{conrad2022emulsion}), one might expect NS condensate to increase the ionic strength in the nanochannel by $\sim75$~mM \ce{NaCl} relative to the homogeneous phase.
That the ionic strength in the nanochannel increases by only $\sim50$~mM~\ce{NaCl} ($\approx35\%\times150$~mM~\ce{NaCl}), suggests that only $\sim 2/3 = 66\%$ of the counterions within the condensate are mobile. 

Regarding the speed with which NS condensate filled the channel: 
Given that four-armed NS condensate is a non-Newtonian fluid with a yield-strain $\gamma_m \sim 0.5$~\cite{conradPNAS_2019}, its flow through the nanochannel is likely plug-like.
The speed with which NS condensate filled the channel, $\mathrm{v}_\mathrm{NSc}$, can therefore be interpreted as the amount by which the EPV of NS condensate exceeds the EOF.
Taking the EOF of buffer alone (PBS150) as an upper bound yields:  
\begin{align*}
30\,\mu\mathrm{m/s} \approx \mathrm{v}_\mathrm{NSc} \leq \,\mathrm{EPV}_\mathrm{NSc} \leq \mathrm{v}_\mathrm{NSc} + \mathrm{EOF}_\mathrm{PBS150} \approx 80\,\mu\mathrm{m/s}
\end{align*}
which constrains the electrophoretic mobility of NS condensate:
\begin{equation*}
    7.5\cdot10^3\mu\mathrm{m}^2/\mathrm{V}\cdot\mathrm{s} = \frac{30\,\mu\mathrm{m/s}}{4\,\mathrm{V/mm}} \lesssim \mu_\mathrm{NSc} \lesssim \frac{80\,\mu\mathrm{m/s}}{4~\mathrm{V/mm}} = 20\cdot10^3 \mu\mathrm{m}^2/\mathrm{V}\cdot\mathrm{s}
\end{equation*}
or between 26\% and 66\% of that of a single NS.

Finally, and perhaps most significantly for sensing applications, the slight variation in $\alpha$-channel current consistently observed upon cooling when EPV is directed uphill (Fig.~\ref{Figure measurement}, inset) suggests that reliable electronic detection is possible even when little or no NS condensate flows through the nanochannel.

\subsection{\label{zetaPotentialMeasurement}Effect of NS and NS condensate on $\zeta$-potential} 
To understand this slight temperature-dependent variation in $\alpha$-channel current that consistently appears while cooling when EPV is directed uphill,
we first consider how exposure to NS and NS condensate alter the EOF in the nanochannel.
Both single-stranded and double-stranded DNA are known to stick to silica surfaces, especially in salt solutions~\cite{Shi_DNA_SiO2_2015}.
DNA adsorption can change the electrical potential at the shear plane adjacent to the nanochannel surface, also known as the $\zeta$-potential, which determines the EOF in the channel (Supp Info, section~2).
We determined the $\zeta$-potential from current monitoring in nanochannels filled with our three different PBS solutions (without DNA).
We did this both before and after subjecting tilted nanochannels to two rounds of heating and cooling under the conditions that created the $\alpha$-channel current signal in question (50~$\mu$M NS, 4~V/mm directing EPV uphill).  

Before exposure, the $\zeta$-potential ranged from -12~mV in PBS300 to -27~mV in PBS050 (Table~\ref{tbl:zeta}).
After exposure, in the case of $\tau$-NS, these values barely changed ($\leq6$\%), independent of the salt concentration, as expected~\cite{crisalli2015label}.
Similarly, in the case of $\alpha$-NS in PBS050, the change in $\zeta$-potential, if any, was barely detectable. 
However, following exposure to $\alpha$-NS in higher salt buffers (PBS150 and PBS300), the magnitude of the $\zeta$-potential increased significantly ($\geq40$\%). 
All $\zeta$-potentials returned to pre-exposure levels after cleaning.

\begin{table}[tbp]     
\centering
\setlength{\tabcolsep}{4pt}     
\setlength{\cmidrulekern}{0.25em} 
        \caption{$\zeta$-potential in buffer-filled nanochannels at $20^\circ$C before and after exposure to 50~$\mu$M~NS in different [\ce{NaCl}]. Uncertainties are standard deviations of three trials}
      \begin{tabular}{
      l
      S[table-format=3.1]@{\,\(\pm\)}S[table-format=1.1]
      S[table-format=3.1]@{\,\( \pm \)\,}S[table-format=1.1]
      S[table-format=1.1]@{\,\( \pm \)}S[table-format=0.1]
      S[table-format=3.1]@{\,\( \pm \)}S[table-format=0.1]
      S[table-format=3.1]@{\,\( \pm \)\,}S[table-format=0.1]
      S[table-format=1.2]@{\,\( \pm \)\,}S[table-format=1.2]       
}
        \toprule
        &\multicolumn{6}{c}{$\alpha$-NS} &\multicolumn{6}{c}{$\tau$-NS}\\
         \cmidrule(rlrlrl){2-7}
         \cmidrule(rlrlrl){8-13}
       &\multicolumn{2}{c}{$\zeta_\mathrm{before}$ (mV)} & 
        \multicolumn{2}{c}{$\zeta_\mathrm{after}$ (mV)} & 
        \multicolumn{2}{c}{$\zeta_\mathrm{after/before}$} & 
        \multicolumn{2}{c}{$\zeta_\mathrm{before}$ (mV)} & 
        \multicolumn{2}{c}{$\zeta_\mathrm{after}$ (mV)} & 
        \multicolumn{2}{c}{$\zeta_\mathrm{after/before}$}\\
\midrule
         {PBS300 } & -12.2 & 0.3 & -21.6 & 1.7 & {\color{black}1.8} & 0.1 & -12.2 & 0.2 & -13.0 & 0.1 & 1.06 & 0.02\\
         {PBS150 } & -18.6 & 0.2 & -26.9 & 3.5 & {\color{black}1.4} & 0.2 & -18.0 & 0.7 & -17.5 & 0.2 & 0.97 & 0.04\\
         {PBS050 } & -25.4 & 2.1 & -28.8 & 1.3 & 1.1 & 0.1& -27.0 & 0.6 & -27.2 & 0.4 & 1.01 & 0.03\\
        \bottomrule
      \end{tabular}
      \label{tbl:zeta}
\end{table}

The unchanging $\zeta$-potential of the $\tau$-channel indicates that {\em uncondensed} NS bind silica weakly, if at all, even when electrostatic repulsion is well-screened.
The more negative $\zeta$-potential of the $\alpha$-channel following exposure to NS and NS condensate in the two higher ionic strength buffers indicates that some NS condensate entered the nanochannel and adsorbed to the silica surfaces.
This suggests that the multiplicity of single-stranded nucleic acid sites on the surface of NS condensate creates multi-valency that stabilizes its interactions with the silica surface.
That the $\zeta$-potential in higher ionic strength buffers rose to levels comparable to the $\zeta$-potential in PBS050, which itself changed only a little, further suggests that the very negative $\zeta$-potential in PBS050 severely limits the amount of NS condensate that enters the nanochannel and adsorbs to its surfaces.

Note that these $\zeta$-potentials were measured after thoroughly rinsing and filling the nanochannels with NS-free buffer.
The presence of NS and NS condensate in the buffer would likely support a more substantial surface coating and, consequently, a more negative $\zeta$-potentials and faster EOF.
The $\zeta_\mathrm{after}$ in Table~\ref{tbl:zeta} therefore represent a lower bound on the $\zeta$-potentials in a nanochannel filled with the same buffer containing NS and NS condensate.
 
\subsection{Electronic detection of condensation}
With these results in mind, we now investigate in detail the slight temperature-dependent variation in $\alpha$-channel current that occurs in a tilted nanochannel when EPV is directed uphill while sweeping temperature ($\pm0.5^\circ$C/min) and holding for 15-min between sweeps.
We focus on this signal because it requires significantly less time and material than filling the nanochannel with NS condensate, and so is better suited to technological application and development.
To improve signal-to-noise and remove the effect of the changing viscosity of the aqueous solution, we look at the ratio of current in the $\alpha$-channel to current in the $\tau$-channel, $I_{\alpha/\tau}(T)$.
We performed three trials of two complete temperature sweeps (hot $\rightarrow$ cold $\rightarrow$ hot $\rightarrow$ cold $\rightarrow$ hot) for each of three NS concentrations, [NS]~=~\{50, 100, 150\}~$\mu$M in each of our three buffers, with nominal salt concentrations [\ce{NaCl}]~=~\{50, 150, 300\}~mM.

\begin{figure}
    \begin{center}
        \includegraphics[width=10cm]{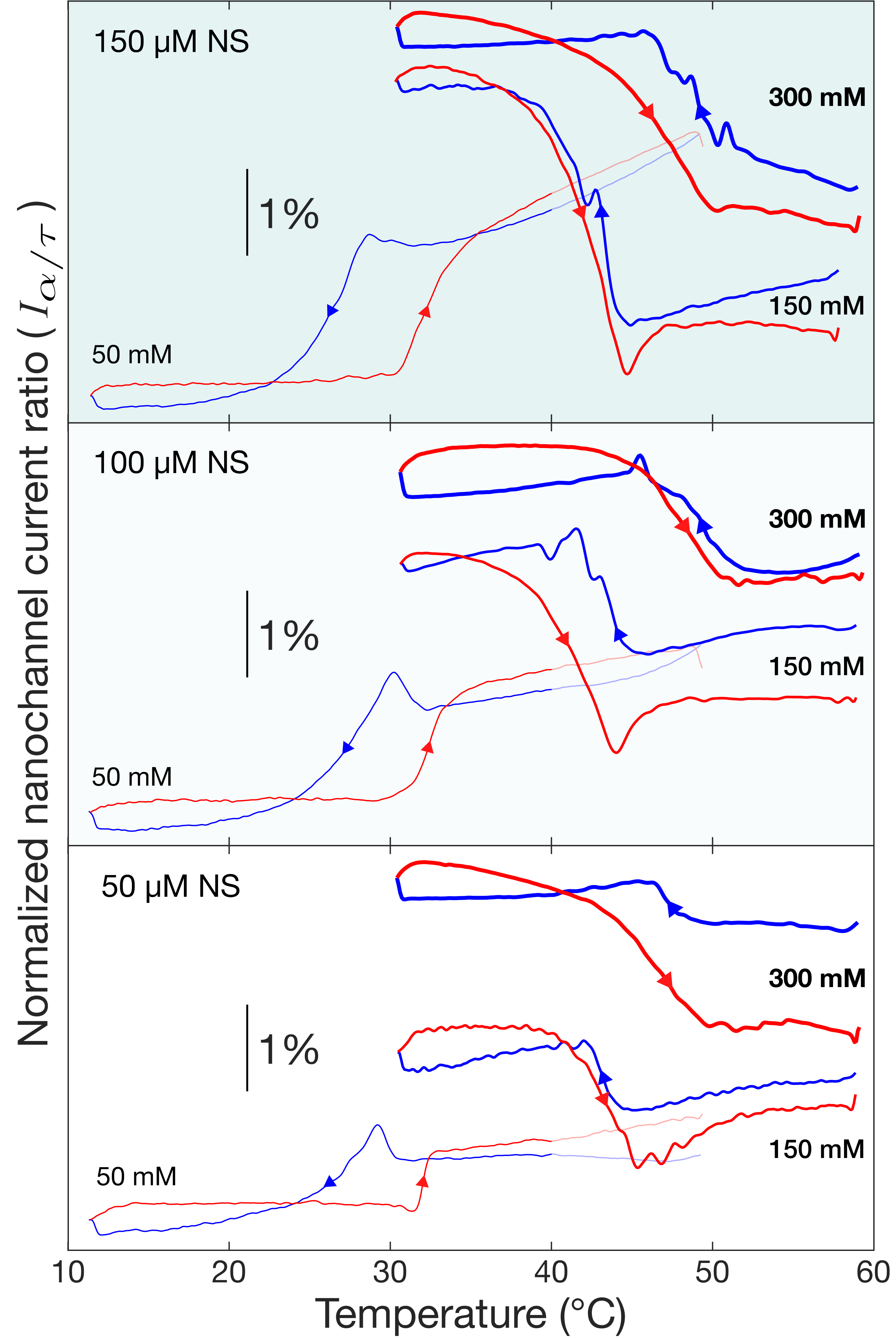}
    \end{center}
\caption{
    Typical current ratio curves, $I_{\alpha/\tau}(T)$, normalized to their low-temperature values.
    (Note that the normalization points have been shifted in the vertical direction to accommodate three curves within each sub-graph.)
    Key features include the opposite sign of the current change in 50~mM~\ce{NaCl}, and the non-monotonicity of the heating curve in 150~mM~\ce{NaCl} and of the cooling curve in 50~mM~\ce{NaCl}.
    See text for a detailed discussion.
}
  \label{fgr:ratio}
\end{figure}

In every case, cooling caused an abrupt change in $\alpha$-channel current that reversed upon heating (Fig.~\ref{fgr:ratio}).
The magnitude of the change increased with [NS], but was never more than a few percent; the effect of [\ce{NaCl}] was more striking.
In $300$~mM~\ce{NaCl}, for all [NS], $I_{\alpha/\tau}$ rose upon cooling and fell upon heating.
In $150$~mM~\ce{NaCl}, for all [NS], $I_{\alpha/\tau}$ behaved similarly except the rise and fall occurred at lower temperatures and the heating curve was distinctly non-monotonic: the decreasing current ratio consistently overshot, dipping and rising before reprising the high temperature behavior of the cooling curve.
In $50$~mM~\ce{NaCl}, again for all [NS], the changes in $I_{\alpha/\tau}$ occurred at even lower temperatures but were opposite in sign: upon cooling, it fell; upon heating, it rose.
The fall was much more gradual than the rise, giving the curves a hysteretic appearance.
Furthermore, reminiscent of the heating curve in $150$~mM~\ce{NaCl}, the cooling curve in $50$~mM~\ce{NaCl} was non-monotonic: the decrease in $I_{\alpha/\tau}$ upon cooling was invariably preceded by a slight rise. 

These observations raise many questions:
\begin{itemize}
    \item What limits the change in $I_{\alpha/\tau}$ to only a few percent? 
    \item Why is the sign of the change different at lower salt?
    \item What causes the hysteresis seen at low salt?
    \item Which features of $I_{\alpha/\tau}(T)$, if any, provide a reliable measure of the transition temperature for NS condensation, $T_c$?
\end{itemize}
The higher conductivity of NS condensate and its affinity for the nanochannel surfaces together suggest testable hypotheses answering these questions.

We propose that $I_{\alpha/\tau}$ increases when NS condensate either forms in or enters the nanochannel.
We further note that while cooling at the two higher salt concentrations (150~mM and 300~mM), the increase in $I_{\alpha/\tau}$ is often rugged.
Given that our setup enables NS condensate to accumulate in the upper reservoir without entering the channel (Subsection~\ref{ionicStrengthMeasurement}), we hypothesize that 
({\em i}) $I_{\alpha/\tau}$ begins rising when NS condensate forms within the $\alpha$-channel because NS condensate has a higher concentration of counterions than the bulk; and, 
({\em ii}) bumps in the rising $I_{\alpha/\tau}$ are the result of large, discrete amounts (droplets) of NS condensate that formed in the lower reservoir entering and passing through, or adsorbing onto, the nanochannel.
Then, $I_{\alpha/\tau}$ stops rising after increasing by only a few percent.
Given that this is far less than the 35\% increase that corresponds to filling the $\alpha$-channel with NS condensate (Subsection~\ref{ionicStrengthMeasurement}), and that exposure to NS condensate augments the $\alpha$-channel $\zeta$-potential (Subsection~\ref{zetaPotentialMeasurement}), 
we hypothesize that $I_{\alpha/\tau}$ stops rising because NS condensate adsorbs to the nanochannel surface, increasing EOF until it equals EPV in magnitude.

At that point, NS condensate is neither drawn into nor pushed out of the $\alpha$-channel.
After it stops increasing, $I_{\alpha/\tau}$ decreases, albeit very slowly. 
This might result from unadsorbed NS condensate drifting out of the $\alpha$-channel, or it might reflect the decreasing ionic strength of the dilute phase, which occurs as NS condensate continues to accumulate in the reservoirs, sequestering counterions. 

At the lowest salt concentration (50~mM), the increase in $I_{\alpha/\tau}$ upon cooling is smooth, short-lived and abruptly reversed.
Its smoothness indicates that large droplets do not enter.
Its brevity suggests that the $\zeta$-potential in 50~mM \ce{NaCl} is so large that only a little NS condensate adsorption suffices to increase EOF to the point that NS condensate is no longer drawn into the nanochannel. 
The abrupt reversal in $I_{\alpha/\tau}$ is harder to explain.
Even if all the NS condensate that forms in or enters the $\alpha$-channel exits without adsorbing, it would not cause $I_{\alpha/\tau}$ to fall below its high temperature value.

We hypothesize that the deep, decelerating decline in $I_{\alpha/\tau}$ results from an asymmetric distribution of adsorbed NS condensate {\color{black}(Supp Info, Section 4)}.
In an electric potential gradient, as counter-ions are drawn through a charged, conductive material and co-ions are drawn away from it, an ion-depleted region forms -- a phenomenon known as concentration polarization (CP)~\cite{SantiagoCP-experiment,SantiagoCP-theory,YangCP}.
Given that adsorbed NS condensate is a negatively charged material with the conductivity of $\sim200$~mM NaCl (Subsection~\ref{ionicStrengthMeasurement}), the CP region that forms in PBS050 may significantly reduce the net current. 
If adsorbed condensate is asymmetrically distributed along the nanochannel, the depletion region may even expand along the length of the nanochannel, progressively reducing its overall conductance~\cite{Sumita_nf_Diodes}. 
Once a 
depletion region is established, if the electrostatic driving potential stays fixed, it cannot be replenished.
Only a complete loss of the perm-selective condensate can eliminate CP~\cite{permSelective}.
In PBS050, this occurs abruptly just above the temperature at which $\alpha$-channel conductivity first began to rise upon cooling (Fig.~\ref{fgr:ratio}, 50~mM cases).


Under all conditions, heating returns $I_{\alpha/\tau}$ to its starting value, modulo a difference we attribute to subtly different rates of evaporation from the different reservoirs.
In 300~mM and 150~mM salt, the current ratio undergoes a smooth decline as NS condensate inside the channel ``dissolves'' and the NS and counterions it contained move into the reservoirs.
Often the decline begins immediately upon heating and is steepest at a temperature well below the one at which it rose upon cooling.
We hypothesize that this asymmetry results because any loss of adsorbed NS condensate reduces the EOF, increasing the speed (EPV - EOF) at which condensate exits the nanochannel. 
Dissolution and clearing may happen faster in PBS150 than in PBS300 because the NS condensate in PBS150 is less stable, having incubated at a temperature ($30^\circ$C) closer to its transition temperature.  
This would explain the dip in $I_{\alpha/\tau}(T)$ when heating in PBS150 as a consequence of the $\alpha$-channel emptying before all the NS condensate in its reservoirs has dissolved. 
The subsequent rise occurs because the $\alpha$-channel contains only dilute phase, which increases in ionic strength until the reservoirs contain no more NS condensate and become homogeneous in ion concentrations. 

\begin{figure}[h]
\includegraphics[width=15cm]{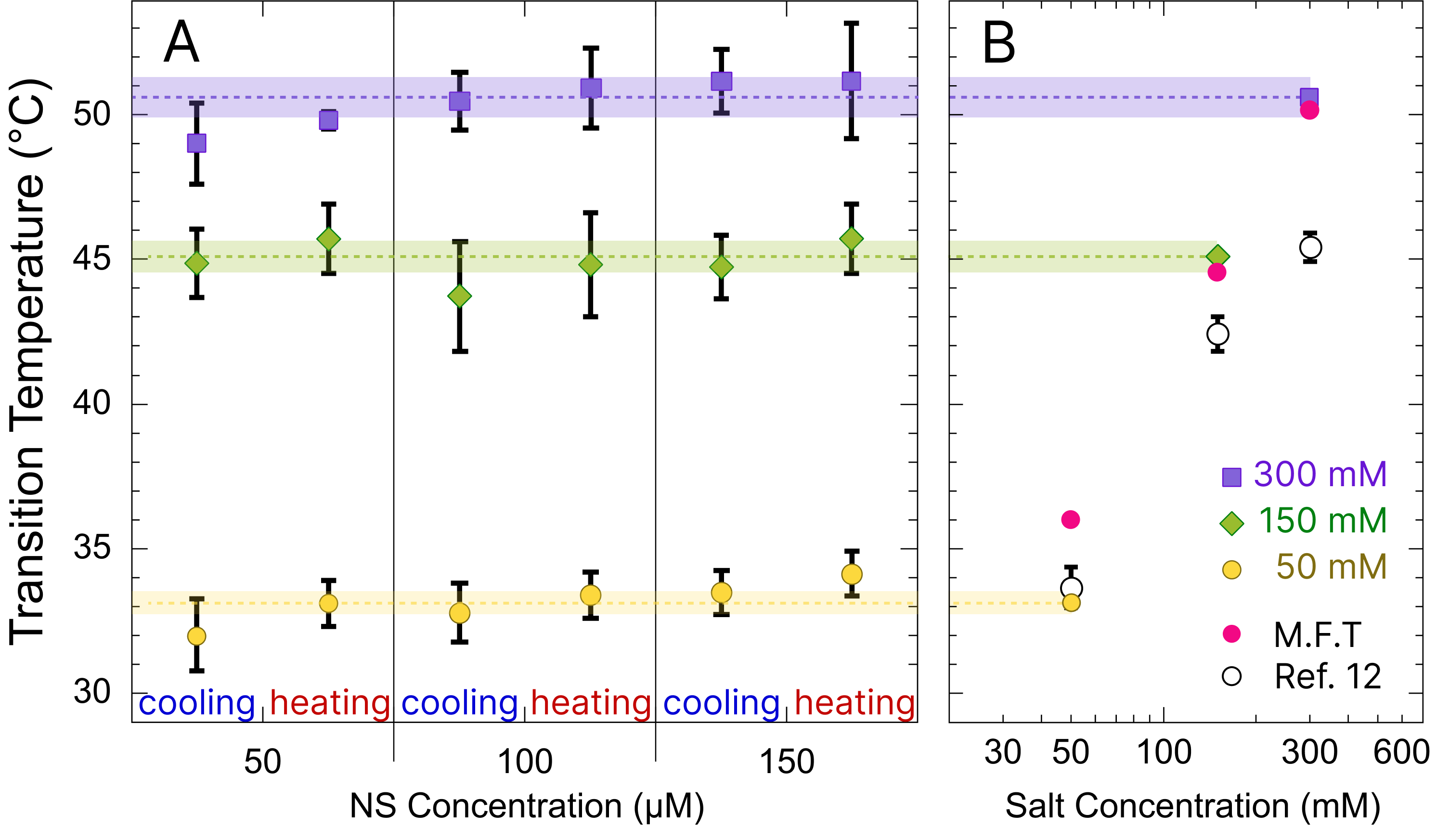}
\caption{
    {\bf (A)} Average (n~=~6) and standard deviation of the NS condensate transition temperature, $T_c$, defined as the highest temperature at which $I_{\alpha/\tau}$ deviates from its high temperature behavior. $\langle T_c\rangle$ increases with [\ce{NaCl}] but is insensitive to [NS].  
    {\bf (B)} The salt concentration dependence of $\langle T_c\rangle$ is similar to that previously observed by direct imaging in 50~mM~\ce{NaCl} but more consistent with a mean-field model in 150~mM and 300~mM~\ce{NaCl}. 
}
  \label{fgr:TCvDNAvNaCl}
\end{figure}

Based on these insights and hypotheses, we associate the transition temperature for NS condensation, $T_c$, with the highest temperature at which $I_{\alpha/\tau}$ deviates from its high temperature behavior (Supp Info, Section 5).
The averages of this $T_c$ over six measurements (three trials, two temperature sweeps each) at each condition are summarized in Fig.~\ref{fgr:TCvDNAvNaCl}A. 
The standard deviation in $T_c$ is as large as $\pm 2^\circ$C under some conditions, but the mean, $\langle T_c\rangle$, estimated by averaging over all [NS], has a standard error of $< 1^\circ$C at the higher [NaCl] and $< 0.5^\circ$C at the lowest.

Interestingly, these $\langle T_c\rangle$ were consistent with measurements made by fluorescence microscopy~\cite{conrad2022emulsion} only at the lowest [\ce{NaCl}] (Fig.~\ref{fgr:TCvDNAvNaCl}B).
At the higher [\ce{NaCl}], they were in better agreement with the predictions of a mean-field model~\cite{conrad2022emulsion} based on the thermodynamics of $\alpha$-NS overhang hybridization~\cite{santalucia2004thermodynamics}.
{\color{black}We suspect that the discrepancy between the electrical and microscopy-based measurements of $T_c$ at higher [\ce{NaCl}] reflects limitations of the latter.  

In reporting their microscopy-based measurements of $T_c$, the authors note that “it was difficult to identify a single frame in which the condensate appeared or disappeared”.  
$T_c$ was therefore estimated as the average $T$ over a range of frames: from the last frame in which a phase boundary was clearly present to the first frame in which it was clearly absent (upon heating); and, from the last frame in which a phase boundary was clearly absent to the first frame in which it was clearly present (upon cooling).

This approach is plausibly biased to underestimate $T_c$ in high [\ce{NaCl}]. 
First, consider cooling.  
At high salt concentration, the barrier to condensate nucleation is low, so upon reaching $T_c$, many sub-optical droplets of condensate rapidly form and deplete the dilute phase [NS].  
Those sub-optical droplets must then grow by coalescing rather than by adsorbing single NS.  
Because small droplets diffuse much more slowly than single NS and take longer to sediment than large droplets, a distinct phase boundary forms only slowly, resulting in a lower measured $T_c$. 
Next, consider heating.
For any given total [NS] and temperature, high salt concentrations result in smaller volume fractions of denser condensate than low salt concentrations \cite{conrad2022emulsion}.  
The larger number of NS released at the surface of a denser condensate may obscure the phase boundary before it has actually disappeared, again resulting in a lower measured $T_c$. 

Confidence in the electrical measurement of $T_c$, is reinforced by having measured the same value at two different, well-controlled rates $|dT/dt| = \{0.5^\circ\mathrm{C/min,}\,0.25^\circ\mathrm{C/min}\}$ of both heating and cooling (Fig. S8).  
The rates of heating and cooling used during optical measurements were different, both from one another and at the different salt concentrations (Ref. 12, Fig S6).}

Finally, we note that, while there exists a nucleation barrier to condensation~\cite{shimobayashi2021nucleation},
we observed no significant change in $T_c|_\mathrm{heat}-T_c|_\mathrm{cool}$ over the three-fold increase in [NS] in any buffer, suggesting that heterogeneous nucleation was the dominant process in our experiments.


\newpage
\section{Conclusion}
We have demonstrated reliable detection of NS condensate using electrokinetic nanofluidics.  
Our approach was enabled by several physical properties of NS condensate:
\begin{itemize}
    \item \textbf{High density of mobile ions} The high linear density of phosphates causes counter-ions to condense along a dsDNA backbone. Because NS condensate is, essentially, a percolated network of dsDNA, when NS condensate fills a nanochannel, a large proportion of the condensed counter-ions are free to move throughout the network, increasing electrical conductivity within the channel.

    \item \textbf{Multiple DNA binding sites} The surface of NS condensate contains numerous DNA binding sites, resulting in multivalent interactions with the silica surfaces of the nanochannel. This allows NS condensate to bind to the silica surface, making the $\zeta$-potential more negative and causing a faster electroosmotic flow (EOF) that opposes the electrophoretic velocity (EPV) of both NS and NS condensate.
 
    \item \textbf{Increased mass density} The mass density of NS condensate is higher than that of the surrounding dilute phase, leading to sedimentation.   As a result, a gentle tilt can accumulate NS condensate near one nanochannel entrance and away from the other, allowing voltage polarity to control whether NS condensate is largely prevented from entering or completely fills the nanochannel.
    
\end{itemize}
A subtle yet characteristic increase in nanochannel current occurs upon LLPS in a tilted nanochannel filled with high ionic strength (150~mM or 300~mM~\ce{NaCl}) solution. 
We explain this as a consequence of the changing balance between uphill EPV and downhill EOF that occurs as NS condensate sticks to the nanochannel walls.
An equally characteristic but more dramatic decrease in nanochannel current occurs following LLPS in a tilted nanochannel filled with low ionic strength (50~mM~\ce{NaCl}) solution. 
We explain this as a consequence of the perm-selective nature of NS condensate, which asymmetrically coats the nanochannel surface coating and mediates ion concentration polarization.

These explanations require further validation, and the assays that inspired them require further development ({\em e.g.,} to increase the speed and reduce the amount of NS required). 
Nevertheless, based on our findings, nanofluidic electrokinetics offers a promising, new approach towards robust, high-throughput detection and characterization of NS condensate and LLPS, and their biomanufacturing, biosensing, and material science applications.

\begin{acknowledgement}

The authors thank Z. Espley and B. Nguyen for preparing nanostar stock solutions and L. Zhou and P.W.K. Rothemund for helpful discussions.  This work was supported by NSF FMRG:Bio award number 2134772.  Some supplies were provided by the Sloan Foundation via award G-2021-16831.

\end{acknowledgement}





\bibliography{reference}


\section*{Supplementary material}

**Supplemental Information** for
"Electrokinetic nanofluidic sensing of DNA nanostar condensate"
by Chou et al.

# 1   DNA nanostar information

**Sequences:** Each nanostar (NS) was a four-way immobile branched junction assembled from four 49-nucleotide-long synthetic DNA oligomers (Table S1). Every arm of the NS consisted of 20 basepairs of duplex DNA and 9 bases of single-stranded DNA (two at the junction, six in the 3′-overhang at the end of the arm and one between the overhang and the arm) (see Figure 1 in the main text). The $\alpha$-sequences and $\tau$-sequences were identical except in their 3′-overhangs. The 6-base sequence, 5′–CGATCG-3′, at the 3′–end of each $\alpha$-NS arm was self-complimentary, allowing any individual NS to bind with others. By contrast, the 6-base sequence 5′–TTTTTT-3′ at the 3′–end of each $\tau$-NS arm has little to no ability to bind itself or others, making it an ideal reference against which to detect condensation among the $\alpha$-NS.

Table S1: DNA oligomer sequences of $\alpha$- and $\tau$-NS. Nomenclature is derived from prior works.[2,3]

| Name | Sequence | | | $\epsilon$ (mL/mole·cm) |
|------|----------|---|---|------|
| V1α | 5′– CTACTATGGCGGGTGATAAA | AA | CGGGAAGAGCATGCCCATCC A CGATCG –3′ | 484.1 |
| V2α | 5′– GGATGGGCATGCTCTTCCCG | AA | CTCAACTGCCTGGTGATACG A CGATCG –3′ | 458.9 |
| V3α | 5′– CGTATCACCAGGCAGTTGAG | AA | CATGCGAGGGTCCAATACCG A CGATCG –3′ | 483.3 |
| S4α | 5′– CGGTATTGGACCCTCGCATG | AA | TTTATCACCCGCCATAGTAG A CGATCG –3′ | 469.0 |
| V1τ | 5′– CTACTATGGCGGGTGATAAA | AA | CGGGAAGAGCATGCCCATCC A TTTTTT –3′ | 476.4 |
| V2τ | 5′– GGATGGGCATGCTCTTCCCG | AA | CTCAACTGCCTGGTGATACG A TTTTTT –3′ | 451.2 |
| V3τ | 5′– CGTATCACCAGGCAGTTGAG | AA | CATGCGAGGGTCCAATACCG A TTTTTT –3′ | 475.6 |
| S4τ | 5′– CGGTATTGGACCCTCGCATG | AA | TTTATCACCCGCCATAGTAG A TTTTTT –3′ | 461.3 |

# 2   Current Monitoring

Current monitoring is a technique used to measure the electroosmotic velocity (EOF) of fluid in a channel by tracking the electrical current in a channel. This measurement can further be used to infer the surface properties of the channel at particular conditions. When the walls of a charged channel are exposed to an electrolyte, counter-ions in the solution are attracted to the surface, forming a thin layer known as the electrical double layer (EDL).[5] The movement of these counter-ions within the EDL generates a force that drags the surrounding fluid in the same direction under an applied axial electric field. In the case where the double layer thickness is much smaller than the channel height, the EOF velocity can be described by the Helmholtz-Smoluchowski equation:

$$u(y) = -\frac{\epsilon \zeta E}{\eta} \qquad (1)$$

where $\epsilon$ is the permittivity, $\zeta$ is the zeta potential, $\eta$ is the fluid viscosity, and $\psi(y)$ is the electrostatic potential. The zeta potential plays a key role in determining the EOF speed, as it depends on the surface charge, solution concentration, and ion valence,[4] and thus measurements of EOF are used to understand how the zeta-potential changes under various conditions and over time.

To perform the measurement, the channel is first filled with a homogeneous electrolyte. One reservoir is then replaced with the same electrolyte solution at a concentration 10% higher than the original. When an electric field is applied, EOF drives the fluid from the higher concentration reservoir to the lower concentration one, increasing the current in the channel. This continues until the higher concentration solution fills the entire channel. Once the current stabilizes, i.e., $\Delta I = 0$, the solution exchange is complete, and the EOF speed can be calculated using the following equation:

$$v_{\mathrm{EOF}} = \frac{t_{\Delta I=0} - t_0}{L} \qquad (2)$$

Where $v_{\text{EOF}}$ is the EOF velocity, $t_{\Delta I=0}$ is the time when $\Delta I = 0$, $t_0$ is the time when the electric field is applied, and $L$ is the length of the channel. Then, the zeta potential can be calculated by the Helmholtz-Smoluchowski equation (Equation 1). An example of the technique in practice is shown in Figure S4.

We conducted current monitoring in our NaCl solutions both before and after DNAns experiments to assess how the NS affects the channel surface (see Section 3.3, main text). Prior to each experiment, the channels were cleaned with 100 mM $H_2SO_4$ at 60°C for 1 hour, followed by rinsing with Milli-Q water at 20°C until the currents stabilized below 100 pA.

Over time, we observed that the current of the buffer solution increased (Figure S5). This behavior aligns with previous findings that repeated rinsing with pure water can etch silica channels.[1] As a result, the baseline current in both the $\alpha$- and $\tau$-channels gradually increased over time. In nanochannels, even minor changes in cross-sectional area can significantly impact current due to Ohm's law.

## 3 Comparison to bulk conductivity measurement

A comparable measurement of the corresponding decrease in dilute phase conductivity might be performed in bulk with a conventional conductivity meter upon separating the phases by, for example, centrifugation. However, conventional conductivity meters require sample volumes $\sim$ 5 mL to achieve precision on the order of $\pm 0.5\%$. For the 35% greater ionic strength of the condensed phase to lower the ionic strength of the dilute phase by 5%, the volume fraction of the condensed phase would have to be 12.5% of the total volume. For $\alpha$-NS, a condensed phase volume fraction of that magnitude occurs at a total NS concentration of $\sim 40\mu M$.[3] The 5 mL of $\sim 40\mu M$ NS solution required for a single bulk conductivity measurement amounts to $\sim$ 200 nmoles of NS, for which the current typical retail cost of synthesis $\sim$ 1000 USD.

## 4 Calculation of Transition temperature

First, the measured current and temperature data were fit with a smoothing spline to reduce noise (Figure S6). Although the $\tau$-NS current was initially intended as a reference for comparing the $\alpha$-NS current, uneven evaporation rates at elevated temperatures between the two channels made this approach unreliable. Instead, the $\alpha$-NS current was normalized to its highest temperature value, and a set of lines with different slopes anchored at its lowest temperature value and enveloping the normalized current (blue line) were generated (Figure S7a). Each of these lines was then subtracted from the normalized $\alpha$-NS current and the result was re-normalized to its highest temperature value (Figure S7b).

At high temperatures for which no condensate was present, changes in the subtracted slopes were gradual. However, when the conductivity changed sharply, a sign of NS condensate formation, a sharp deviation was evident in the curves. The transition temperature was identified as the point where the curves began to deviate from the high-temperature slopes.

To validate the transition temperature, a line was drawn between the transition temperature and the highest current value, then extended to the lowest temperature. The transition temperature was confirmed if the ratio of the $\alpha$-NS current to this line remained within $\pm 0.1\%$ of 1 at high temperatures (Figure S7c).

(A)

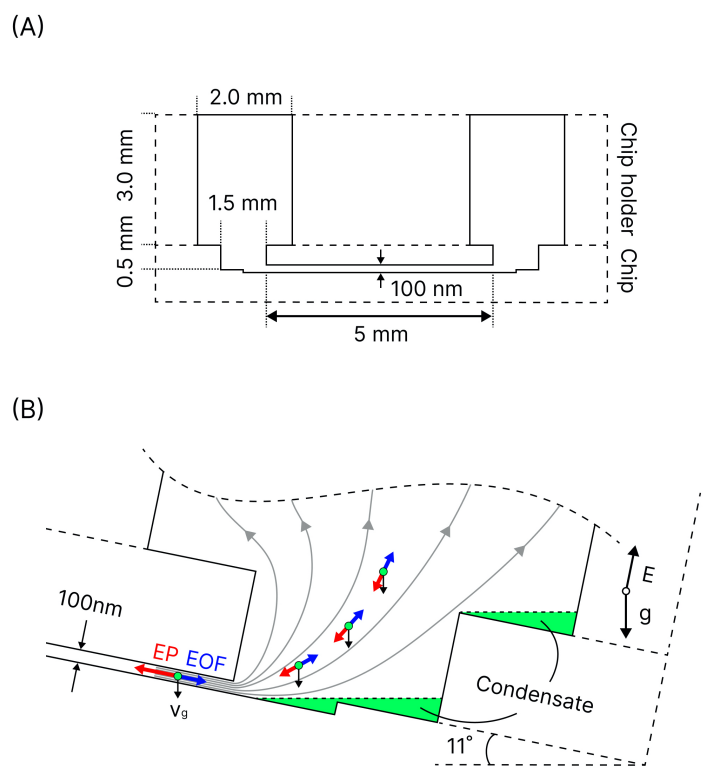

Figure S1: Geometry of the chip in its chip holder **(A)** Cross-section with dimensions. **(B)** Close-up of tilt and its effect on condensate sedimentation.

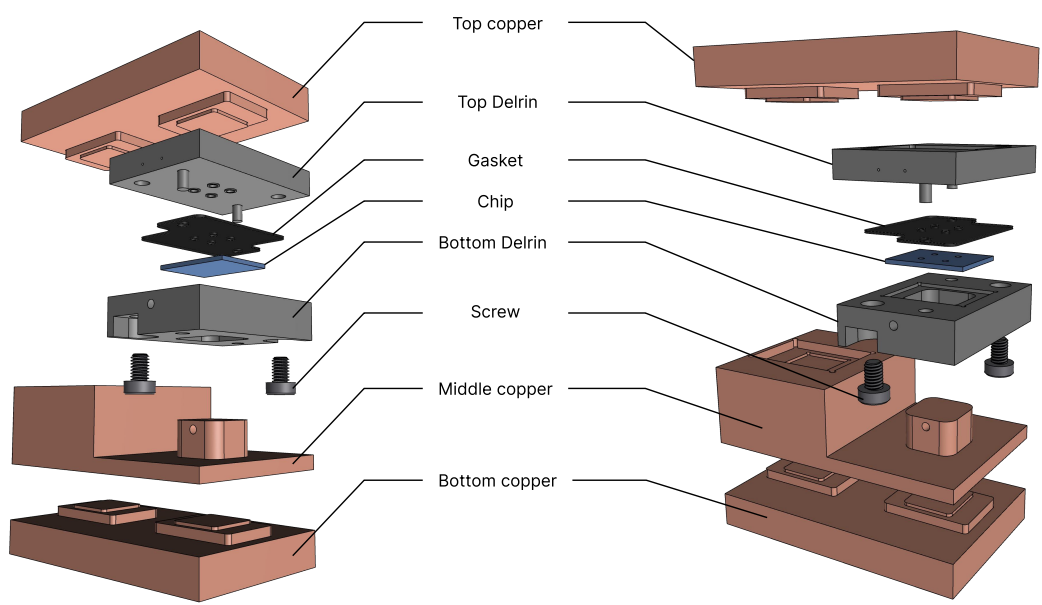

Figure S2: The model of chip holder and the copper components designed to transport heat. **(left)** Bottom view, and **(right)** Top view. The electrode accesses were drilled on the top Delrin. The access for the thermocouple was drilled on the bottom Delrin.

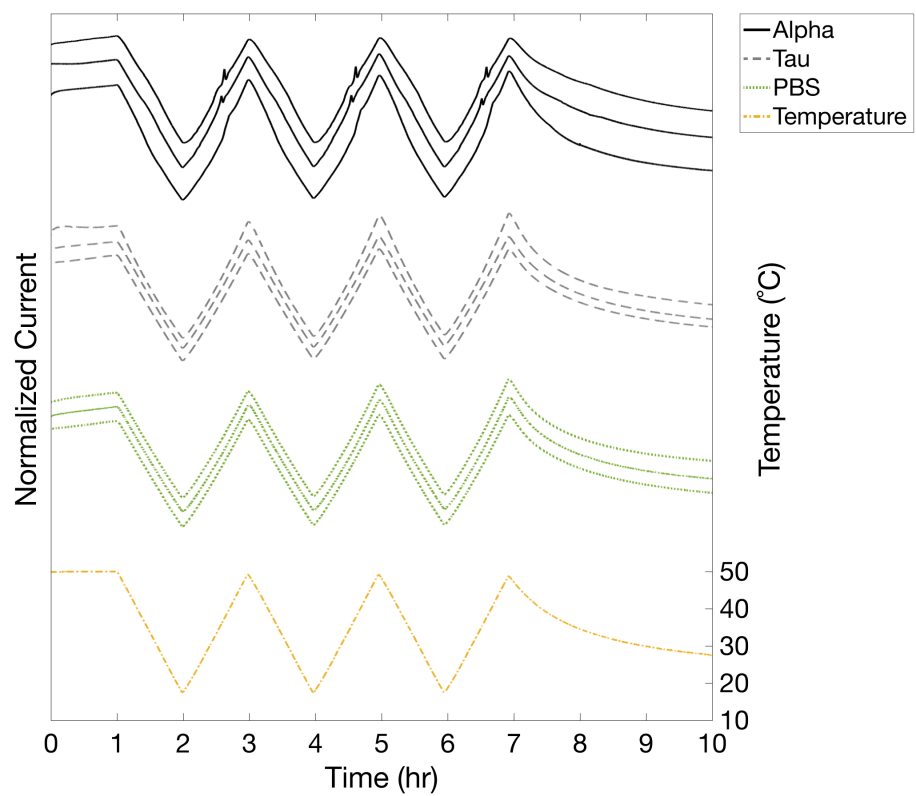

Figure S3: The measurement of NS detection for $\alpha$-, $\tau$-NS, and PBS. An abrupt change occurs only when the channel was filled with $\alpha$-NS.

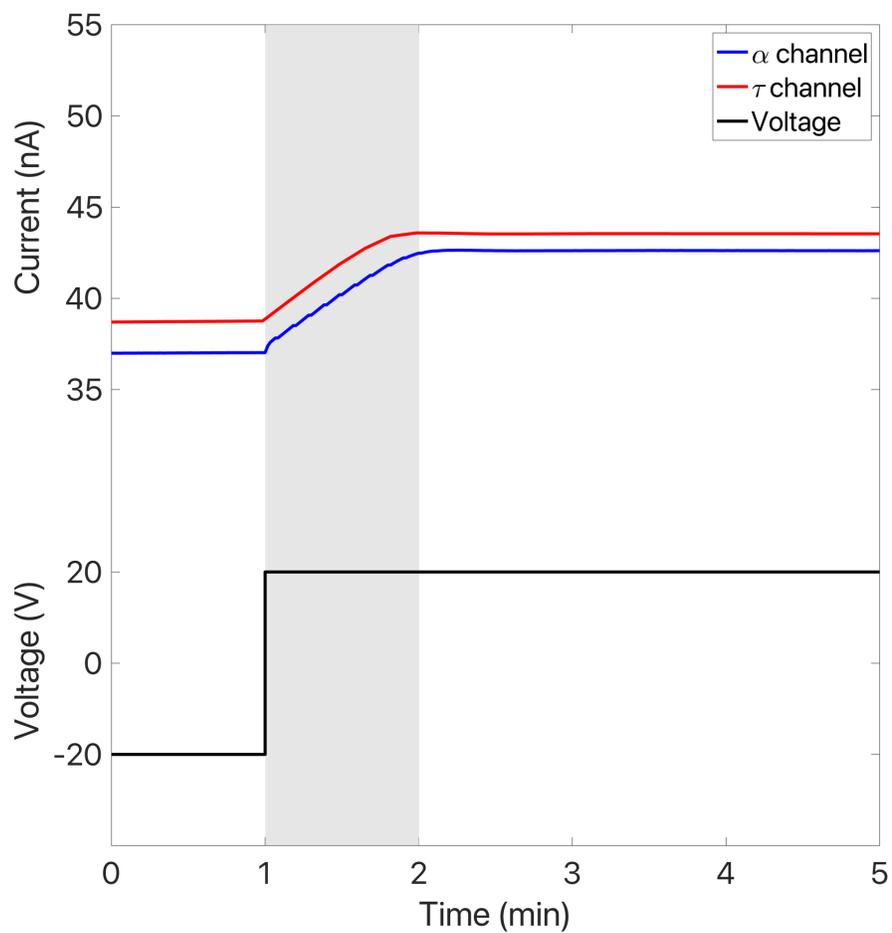

Figure S4: The current monitoring technique. Both channels are filled with 90% of 300 mM NaCl in 19 mM phosphate buffer (PBS300). The positive well with 100% PBS300 and the negative well with 90% PBS300. Before changing the polarity (from -20V to +20V), the current remains at the same level. After changing the polarity, the direction of EOF also changes, causing the currents to continuously increase when the channels are filling 100% PBS300, and to be stable when the channels are fully filled with 100% PBS300.

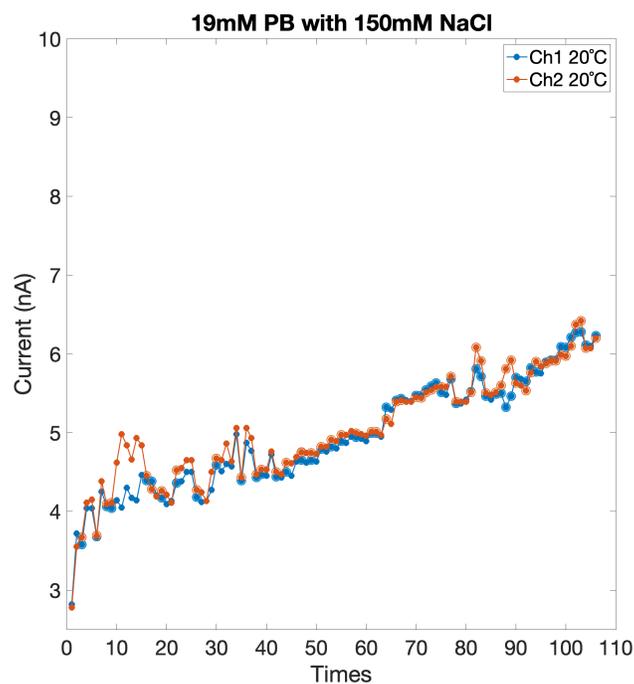

Figure S5: Tracing the current of the buffer solution in the conditioned channel over time

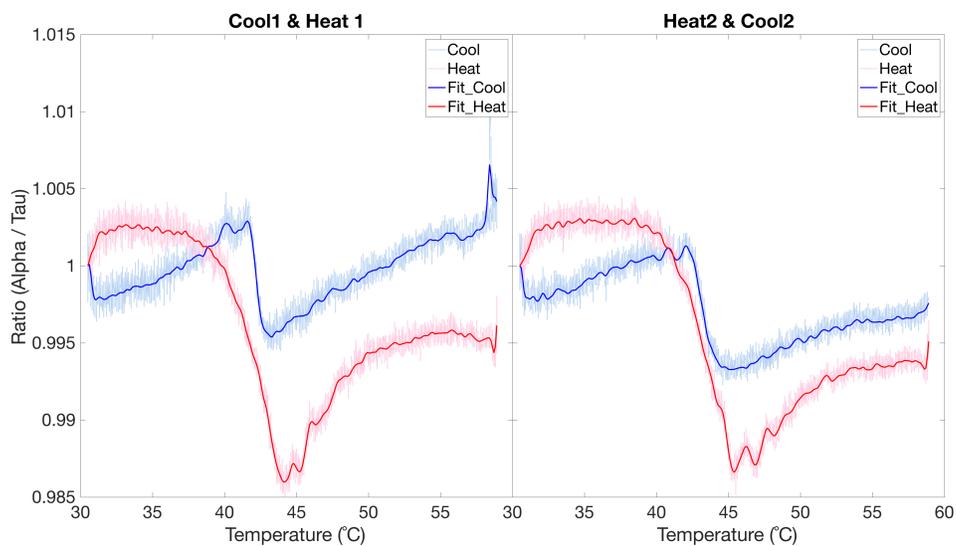

Figure S6: The demonstration of smoothing process for the alpha and tau currents. The light red and blue curves are raw data, and the red and blue are smoothed currents during cooling and heating.

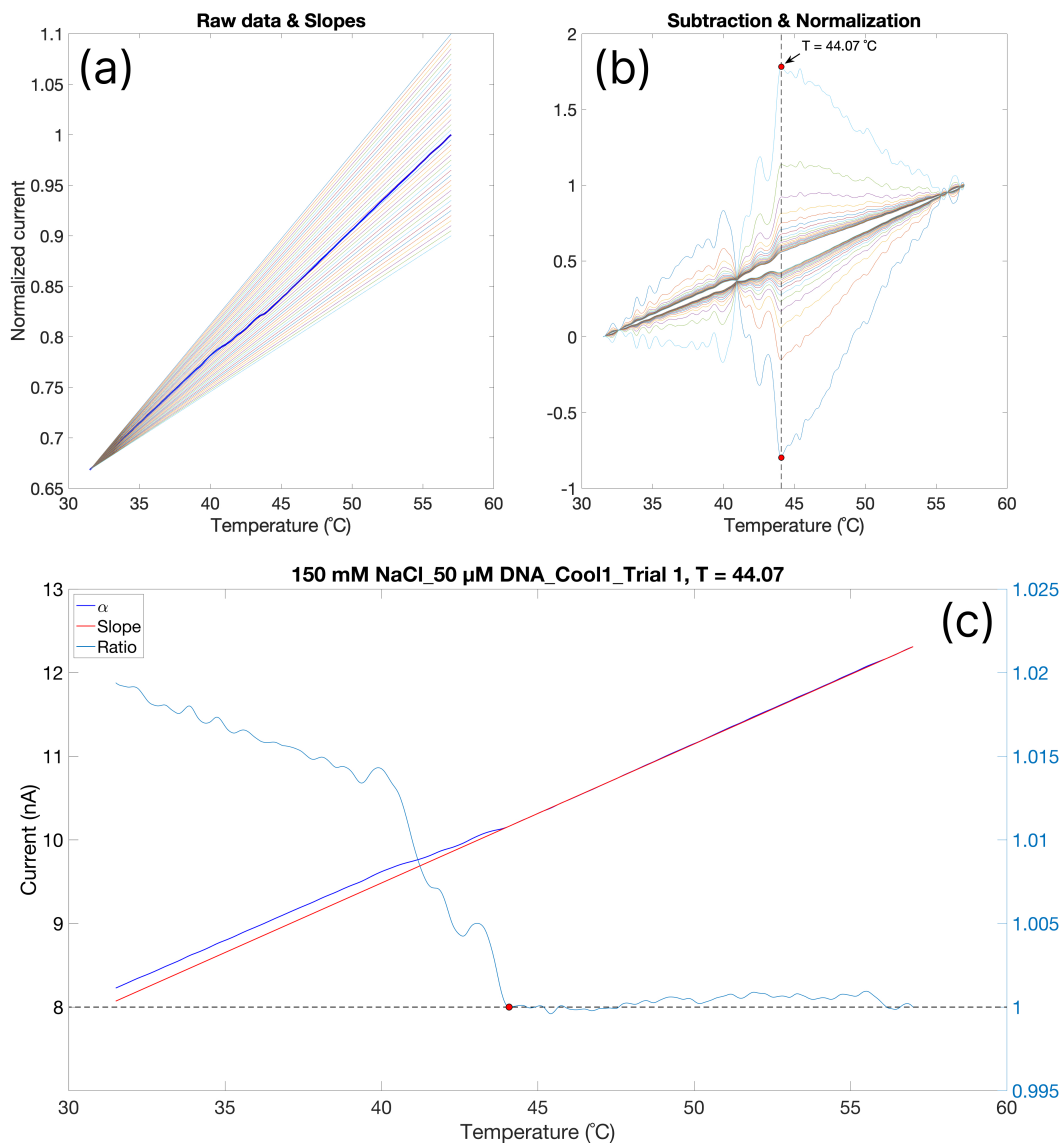

Figure S7: Determination of the transition temperature. (a) The smoothed and normalized $\alpha$-NS current (blue) is used to create a set of lines with varying slopes adjusting the final value at the highest temperature, (b) we subtract the normalized current with each slope generated, and normalized at the highest temperature. We determine the transition temperature to be when the curves deviated significantly, (c) verification of the selected transition temperature is performed by drawing a line with the chosen slope a slope between the current value at the transition temperature and the highest current value and extended to the point at the lowest temperature. The ratio between the current and slope is taken to confirm that the ratio is around 1 in high temperature range.


\end{document}